\let\a=\alpha \let\b=\beta  
\let\d=\delta  \let\e=\varepsilon
 \let\g=\gamma \let\h=\eta   \let\l=\lambda
      \let\o=\omega
  
  \let\s=\sigma \let\t=\tau

  \let\z=\zeta

\let\i=\infty

\def\oo{{\o}}  
\def\xx{{\vec x}} \def\yy{{\vec y}} \def\kk{{\vec k}}

\def\LL{{\cal L}}\def\RR{{\cal R}}
\def\NN{{\cal N}}
\def\DD{{\cal D}}\def\GG{{\cal G}}

\def\nn{\nonumber}

\documentstyle[twocolumn]{article}
\begin{document}


\title{Renormalization group for the XYZ model}

\author{V.Mastropietro\\
        Dipartimento di Matematica, Universit\`a ``Tor Vergata'' di Roma\\
        Via della Ricerca Scientifica, 00133, Roma\\
         tel. 72594688, Fax 72594699, vieri@ipparco.roma1.infn.it}

\maketitle
\begin{abstract}
We study in a rigorous way
the XYZ spin model
by Renormalization Group methods.
\end{abstract}
\vskip.5cm
{\bf pacs 75.10.Jm}
\vskip.5cm
{\bf Renormalization Group, Correlations, XYZ chain }
\vskip.5cm
The $XYZ$ model is exactly solvable [1]
by the {\it transfer matrix formalism}
as it is equivalent [2]
to the eight vertex model. The
solution is so complicated that it is very difficult to compute the
correlation functions from it (an attempt with some
preliminary results is in [3]).
The correlations are then computed
in an approximate way
by linearizing the bands and taking the
continuum limit [4], so introducing spurious u.v. divergences.
One has to introduce an {\it ad hoc}
cut-off, absent in the original model, for applying
the {\it bosonization methods},
and it is not very clear the relationship
of the obtained correlations with the real ones.
Finally {\it Bethe ansatz} and the
{\it conformal algebra} methods cannot be applied
to the $XYZ$ chain but only to its limiting case given by the
$XXZ$ chain [5] (also bosonization results are mainly for this case).

In this letter we apply to the $XYZ$ chain the RG methods developed
for QFT [6]; we show that, for small anisotropy and $J_3$,
the correlations can be expressed by convergent series.
This is almost equivalent to know the correlations
exactly, as one can compute the first orders having a rigorous bound on the
remainder {\it i.e.} the correlations are known up to a small
error. The comparison of the so obtained
correlations with experiments or numerics is then without ambiguity.
No approximation are necessary in our approach, which can be
extended to a variety of models. Bounds on the corrections due to the
finite size effect are naturally obtained, as our results are {\it
uniform} in the size.
We write the $XYZ$ chain as a
system of interacting fermions, and we compute the fermionic
two point different (imaginary) time correlation function.
The computation of the
other correlations is
a straightforward (but cumbersome)
consequence of our results.

If $(S^1_x,S^2_x,S^3_x)=
{1\over 2}(\s^1_x,\s^2_x,\s^3_x)$,
$\s^\a_x$, $\a=1,2,3$ being the Pauli
matrices and $x=1,2,...,N$ the hamiltonian $H$ is
$$\sum_{x=1}^{N}
[J_1 S^1_x S^1_{x+1}+J_2 S^2_x S^2_{x+1}+
J_3 S^3_x S^3_{x+1}+h S^3_x]+U^1_N$$
where the last term is a boundary term and $S^3_{N+1}=S^3_1$ (the
boundary conditions on $S^1_x,S^2_x$ will be specified later fixing
$U^1_N$)).
The {\it anysotropy} is $u={J_1-J_2\over
J_1+J_2}>0$ for fixing ideas.

By a {\it Jordan-Wigner transformation} [7] we write,
if $S_x^{\pm}=S_x^1\pm i S_x^2$,
$S^-_x=e^{-i\pi
\sum_{y=1}^{x-1}\psi^+_y\psi^-_y } \psi^-_x$ and $S^+_x=\psi^+_x e^{i\pi
\sum_{y=1}^{x-1}\psi^+_y\psi^-_y }$, where $\psi^\pm_x$ are fermionic
operators. Moreover $S^3_x=\psi^+_x\psi^-_x-{1\over 2}$.
Fixing  $J_1+J_2=1$, $H$ is:
$$\sum_{x=1}^N\{[\psi^+_{x}\psi^-_{x+1}+\psi_{x+1}^+\psi^-_{x}]
+u[\psi^+_{x}\psi^+_{x+1}+
\psi^-_{x+1}\psi^-_{x}]+$$
$$J_3(\psi^+_x\psi^-_x-{1\over 2})
(\psi^+_{x+1}\psi^-_{x+1}-{1\over 2})+h
(\psi^+_x\psi^-_x-{1\over 2})\}+U_N^2$$
where $U_N^2$ is a boundary term. We choose
$U^1_N$ so that $U^2_N=0$ and the fermions verify periodic boundary
conditions
(in [7] this choice for the XY chain is called
"c-cyclic"). This is achieved by setting $U_N^1=[(-S^+_N
S^-_{N+1}+
S^+_N e^{i\pi\NN}S^-_1)+c.c.]+$ $u[(-S^+_N S^+_{N+1}+S^+_N
e^{i\pi\NN}S_1^-)+c.c.]$, if
$\NN=\sum_{x=1}^N\psi^+_x\psi_x$;
as $[(-1)^\NN,H]=0$ the eigenvectors of $H$
are divided in two subspaces on which $(-1)^\NN$ is equal to $1$ or $-1$
and on such subspaces $U^1_N$ is a boundary term.
Let be
$S^{\e_1,\e_2}S(\xx)=\lim_{N,\b\to\i}
<T \psi^{\e_1}_\xx \psi^{\e_2}_{\vec 0}>_{N,\b}$, $\e_i=\pm$,
$T$ is the time ordering and
$<\cdot>={{\rm tr} e^{-\b H}\cdot\over {\rm tr} e^{-\b H}}$.
We prove that, if $\lim_{N,\b\to\i}S^{\e_1,\e_2}_{N,\b}(\xx)=
S^{\e_1,\e_2}(\xx)$,
$\xx=x,t$ and setting $p_F=
\cos^{-1}(h-J_3)\not= n\pi$, $v_0=\sin p_F$ and
$|\xx|=\sqrt{x^2+v_0^2 x_0^2}$, $\kk=k,k_0$:

{\it For $J_3,u$ suitably small
$$S^{\e_1,\e_2}(\xx)=S_0^{\e_1,\e_2}(\xx)+\e S_1^{\e_1,\e_2}(\xx)$$
where $S_0^{-,+}(\xx)$ and $S_0^{+,+}(\xx)$ are respectively

$$\int d\kk \sin(\bar p_F x+k_0t) {2i e^{i k x}\over Z(k)}
{i k_0-v_0\sin(|k|-\bar p_F)\over k_0^2+v_0^2\sin^2(|k|-\bar p_F)+\s(k)^2}$$
$$\int d\kk \sin(\bar p_Fx) {2ie^{i\kk\xx}\over Z(k)} {\s(k)\over
k_0^2+v_0^2\sin^2(|k|-\bar p_F)+\s(k)^2}$$
where $\e=\max (u,|J_3|,u^{1+\h_2})$,
$\bar p_F=p_F+O(J_3)$ and
$Z(k),\s(k)$ are smooth functions such that
$|Z(k)-1|\le\e$, $|\s(k)-u|\le \e$ for $||k|-\bar p_F|\ge
{\rm min}[(\bar p_F/2,(\pi-\bar p_F)/2]$
and
$$Z(\pm \bar p_F)\equiv \hat Z=u^{-\h_1} \qquad
\s(\pm \bar p_F)\equiv \hat \s=u^{1+\h_2}$$
with
$\h_1=\b_1(J_3)^2+O((J_3)^3),
\h_2=-\b_2J_3+O((J_3)^2)$, $\b_1,\b_2>0$. Moreover
for $|\xx|\ge\hat u^{-1}$,
$S_{0,1}^{\e_1,\e_2}(\xx)$
have a long
distance faster than any power decay, {\it i.e.} for any $M$,
$|S^{\e_1,\e_2}_{0,1}(\xx)|\le {C_M\over\hat Z}
{\hat u \over 1+(\hat u |\xx|)^M}$
if $C_M$ is a constant; for
$1\le |\xx|\le \hat u^{-1}$ they have a transient slow decay
$|S^{\e_1,\e_2}_{0,1}(\xx)|\le {C_1\over |\xx|^{1+\h_3}}$,
with $\h_3=\h_1(1+\h_2)^{-1}$.}

The optimal bound for $S^{\e_1,\e_2}_{0,1}(\xx)$ should be
$|S^{\e_1,\e_2}_{0,1}(\xx)|
\le {c_1 \over
|\xx|^{1+\h_3}}e^{-c_2 u^{1+\h_2}|\xx|}$, $c_1,c_2>0$
constants, and this could be proved
by a slight improvement of our techniques. We see that there is an
{\it anomalous gap} and an {\it anomalous wave function renormalization}.
Moreover the oscillation period $\bar p_F$ depends on $J_3$.
The explicit form of $Z(k)$ and $\s(k)$ will be given below;
if $u=0$  $\s(k)=0$ and $\hat Z(k)=|k|^{-\h_3}$
according with the expected
power law decay of the $XXZ$ chain. A simple consequence of our analysis
is that, in the $J_3=h=0$ case,
$|<S^3_\xx S^3_0>-<S^3_\xx><S^3_{\vec 0}>|\le {c_1 \over
|\xx|^{2}}e^{-c_2 u|\xx|}$ in agreement
with the results in the
$XY$ chain in [7], for small $u$ and $t=0$
$\sin^2({\pi\over 2}x){-1\over 4\pi}({1-u \over 1+u})^{x}{1\over x^2}[1+O({1\over
x})]$.

{\it RG analysis.}
The Grand-Canonical Schwinger
function can be defined by Grassman integrals so we can write
$<T \prod_{i=1}^n \psi^{\e_i}_{\xx_i}>_{L,\b}$ as
\begin{equation}\label{S}
\int\{\DD\psi e^{ -\int d\kk
\psi^+_{\kk}(-ik_0-E(k))\psi^-_{\kk}}\}e^{-V(\psi)}
\prod_{i=1}^n \psi^{\s_i}_{\xx_i}
\over \int\{\DD\psi e^{ -\int d\kk \psi^+_{\kk}(-ik_0-E(k))\psi^-_{\kk}}\}
e^{-V(\psi)}
\end{equation}
where $\kk=2\pi({n_1\over N},
{n_0+2^{-1}\over\b})$,
if $n_0,n_1$ are integers, $E(k)=\cos k-\cos p_F$,
$\psi^\pm _{\xx},
\psi^\pm_{\kk}$,
are {\it Grassman
variables} and we denote by $\int\{\DD\psi e^{
-\int dk \psi^+_{\kk}
h(\kk)^{-1}\psi^-_{\kk}}\}$ the
{\it fermionic integration} with propagator
$\int d\kk e^{i\kk(\xx-\yy)}h(\kk)$,
if $\int d\kk={1\over L\b}\sum_\kk$.
Finally $V(\psi)$ can be written as
$V(\psi)=J_3 \bar V+u P+\nu N$
where
$\bar V=\int \prod_{i=1}^4 d\kk_i \cos(k_1-k_2)
\psi_{\kk_1}^+\psi_{\kk_2}^-\psi_{\kk_3}^+
\psi_{\kk_4}^-\d(\kk_1-\kk_2+\kk_3-\kk_4)$
and $P=\int d\kk [e^{ik}\psi^+_\kk\psi^+_{-\kk}+
e^{-ik}\psi^-_{-\kk}\psi^-_{\kk}]$,
$N=\int d\kk \psi^+_\kk\psi^-_{\kk}$.
Note that $\nu$ is a {\it counterterm}
to be fixed so that the Fermi momentum in the $J_3=0$
or $J_3\not=0$ theory are the same. In fact,
as the oscillation period is
changed by the presence of the $J_3$ term, we find technically convenient
to fix it to its value in the $J_3=0$ case by adding the counterterm.
According to the (formal) Luttinger theorem, this means that
we add a magnetic field so that the mean magnetization
in the direction of the magnetic field is the same as in the $J_3=0$ case.

We start by integrating the denominator of eq.(\ref{S}), the {\it
partition function} $\NN$. We perform a decomposition of the propagator
$g(\kk)=f_1(\kk))g(\kk)+(1-f_1(\kk)) g(\kk)$
where $f_1(\kk)=1-\chi(k-p_F,k_0)-\chi(k+p_F,k_0)$ and
$\chi$ is a smooth compact support function
such that $\chi(k\pm p_F,k_0)$ are non vanishing
only
in two non overlapping regions around $p_F,0$ and $-p_F,0$
respectively.
We call
the two addends respectively $g^{> 0}(\kk)$ and $g^{\leq 0}(\kk)$
and this allows us to represent $\psi_{\kk}^\pm$ as sum of
two independent Grassmanian variables, $\psi_{\kk}^{(>0)\pm}$,
$\psi_{\kk}^{(\leq 0)\pm}$.
The integration on $\psi^{(>0)}$
allows us to write $\NN$ with
$\psi$ replaced by $\psi^{(\le 0)}$ and $V$
replaced by $V^0$ defined as [8]:
\begin{equation}
\sum_{n=0}^\i\int \prod_{i=1}^{n} d\kk_i
W_{n}^h(\kk_1,\ldots,\kk_{n})\prod_{i=1}^{n}\psi^{(\le h)\e_i}_{\kk_i}
\d(\sum_{i=1}^n\e_i \kk_i)\label{c}
\end{equation}
with $h=0$, where
$|W_{n}^0|\leq C^m z^{\max(2,{n}-1)}$
if $z=\max(|\l|,|u|,|\nu|)$. Note that, unless
$u=0$, $\sum_{i=1}^n\e_i\not =0$ as $[H,\NN]\not=0$.
We set $k=k'+\o p_F$, $\o=\pm 1$, ${\rm sign}(\o
k)>0$; moreover we write $1-f_1(\kk)=\sum_{\o=\pm 1}\sum_{h=-\i}^0 f_h(\kk')$,
$f_h(\kk')=\chi(\g^{-h}\kk')-
\chi(\g^{-h+1}\kk')$ and $C_h^{-1}=\sum_{k=-\i}^h f_h$. This allows us
to write $\psi^{(\leq 0)}_{k}=
\sum_{\oo=\pm 1}\sum_{h=-\i}^0 \psi^{(h)}_{k,\oo}$.
In other words we represent the fermions by two Fermi fields
with label $h$,
with momenta close $O(\g^h)$ respectively to $p_F$ or $-p_F$.
We proceed
iteratively setting
$Z_0=1$: once the fields $\psi^{(0)},\ldots,\psi^{(h+1)}$ have been
integrated $\NN$ is given by:
\begin{eqnarray}
&&\int\{ {\cal D}\psi^{(\leq h)}
e^{-\int d\kk' C_hZ_h \vec\psi^{(\leq
h)+}_{\kk'}\GG^{(h)}(\kk')^{-1}
\vec\psi^{(\leq h)-}_{\kk'}}\}\nn\\
&&e^{-V^h(\sqrt{Z_h}\psi^{(\leq h)})}\label{b}
\end{eqnarray}
if $\vec\psi_{\kk'}^{(\le h)\pm}=(\psi^{(\le h)\pm}_{\kk'+p_F,1},
\psi^{(\le h)\mp}_{-\kk'-p_F,-1})$
and $V^h(\psi^{(\le h)})$ is given by terms like eq.(\ref{c}).
If $\a(k)=\cos p_F(1-\cos k)$,
$[\GG^{(h)}(\kk')]^{-1}_{\oo,\oo}=
(-ik_0+\oo v_0\sin\kk +\oo\a(k'))$,
$[\GG^{(h)}(\kk')]_{\oo,-\oo}^{-1}=
i\o \s_h(k')$.
We define a {\it localization operator} $\LL$
extracting the {\it relevant
part}
of the {\it effective potential} $V^h$: i) If $n>4$ then $\LL W^h_n=0$;
ii) if $n=4$
if $\d_4^a=\d_{\sum_{i=1}^4\e_i\o_i
p_F,0}\d_{\sum_{i=1}^4\e_i,0}$, then
$\LL W^h_4(\kk_1+\oo_1 p_F,....)=
\d_4^a W_{4}^{h}(\o_1 p_F,...)$
iii) if $n=2$ then if
$\d_2^a=\d_{(\o_1-\o_2)p_F,0}\d_{\e_1-\e_2,0}$
and $\d_2^b=\d_{(\o_1+\o_2)p_F,0}\d_{\e_1+\e_2,0}$ then
$$\LL W_{2}^{h}(\kk'_1+\o_1 p_F,
\kk'_2+\o_2 p_F)=\d_2^a [W_{2}^{h}(\o_1 p_F,\o_2 p_F)+$$
$$\o_1E(k'+\o_1 p_F)
\partial_{k}
W_{2}^{h}(\o_1 p_F,\o_2 p_F)
+k^0$$
$$\partial_{k_0}W_{2}^{h}(\o_1 p_F,\o_2 p_F)]
+\d_2^b[W_{2}^{h}(\o_1 p_F,\o_2 p_F)]$$
where $E(k'+\o p_F)=v_0\o\sin k'+\a(k')$ (the symbols
$\partial_k,\partial_{k_0}$ means discrete derivatives
and the first deltas in $\d_2^a,\d_2^b,\d_4^b$ are mod. $2\pi$).
A naive
power counting argument explains why the relevant terms are only
the quartic or bilinear in the fields;
moreover among such terms
there are still irrelevant ones {\it i.e.} the power counting can be
improved. This is taken into account in the definition of $\LL$, as
the first of the
two deltas in $\d_4^a,\d_2^a,\d_2^b$
says that the relevant terms involve
only fermions at the Fermi surface, {\it i.e.} $\sum_i\e_i \oo_i
p_F=0$ {\it modulo} $2\pi$,
and the second takes into account
that the marginal terms with $\sum_i\e_i\not=0$ are indeed irrelevant.
This will discussed below.
We can write then the relevant part of the effective potential as:
$$\LL V^{h}=\g^h n_h F_\nu^{h}+s_h F_\s^{h}+z_h F_\z^{
h}+a_h F_\a^{h}+l_h F_\l^{h}$$
where $F_i^{h}=\sum_\o \int d\kk'
 f_i \psi^{(\le h)+\o}_{\o\kk'+\o p_F,\o}
\psi^{(\le h)-\o}_{\o\kk'+\o p_F,\o}$, and
$F_\s^{h}=\sum_\o \int d\kk'
\psi^{(\le h)+\o}_{\o\kk'+\o p_F,\o}
\psi^{(\le h)-\o}_{-\o\kk'-\o p_F,-\o}$ and
$F_\l^{h}$ is given by
$\int [\prod_{i=1}^4 d\kk'_i
\psi^{(\le h)\e_i}_{\kk'_i+p_F,\oo_i}]
\d(\sum_{i=1}^4\oo_i\s_i\kk'_i)$
where $i=\nu,\a,\z$, $f_\nu=\o$, $f_\a=\oo E(k'+p_F)$,
$f_\z=-i k_0$. Moreover
$l_0=J_3+O((J_3)^2)$,
$s_0=u+O(u J_3)$, $a_0,z_0=O(J_3)$,
$n_0=\nu+O(J_3)$ and we have defined $\psi^{{(\le h)\pm \o}}=
\psi^{{(\le 0})\pm}$ if $\o=1$ and $\psi^{{(\le 0)\o\pm}}=
\psi^{{(\le 0)\mp}}$ if $\o=-1$.
We write eq.(3) as:
\begin{eqnarray}\label{neo6}
&&\int {\cal D}\psi^{(\leq h)}
e^{-\int dk' \vec\psi^{(\leq
h)+}_{k'}C_hZ_{h-1}(k')\GG^{(h-1)}(k')^{-1}
\vec\psi^{(\leq h)-}_{k'}}\nn\\
&&e^{-\tilde V^h(\sqrt{Z_h} \psi^{(\leq h)})}]
\end{eqnarray}
where $\tilde V^h=\LL \tilde V^h+(1-\LL) V^h$,
$\LL \tilde V^h=\g^h n_h F_\nu^{h}+
(a_h-z_h) F_\a^{h}
+
l_h F^{h}_\l$
and $Z_{h-1}(k')=Z_{h}+C_h^{-1} Z_h z_h$,
$Z_{h-1}(k')\sigma_{h-1}(k')=Z_h\sigma_h(k') +Z_h C_h^{-1} s_h$.
This means that we extract from the effective potential the terms
leading to a mass and wave function renormalization.
Now one can perform the integration respect to $\psi^{(h)}$
rescaling the effective potential
$\hat V^h(\psi)=\tilde V^h(\sqrt{Z_h\over Z_{h-1}}\psi)$ and
$\LL \hat V^h=\g^h \nu_h F_\nu^{h}+
\d_h F_\a^{h}+
\l_h F^{h}_\l$
with $\g^h\nu_h={Z_h\over Z_{h-1}}n_h$,
$\d_h={Z_h\over Z_{h-1}} (a_h-z_h)$,
$\l_h=({Z_h\over Z_{h-1}})^2 l_h$ and $\vec
v_k=\nu_h,\d_h,\l_h$.
The
integration of $\psi^{(h)}$ has propagator
$g^h_{\oo,\oo'}(x-y)={1\over Z_{h-1}}
\int d\kk' e^{i\kk'(x-y)}\tilde f_h(\kk')
\GG^{(h-1)}(\kk')_{\oo,\oo'}$,
if $Z_{h-1}\equiv Z_{h-1}(0)$ and $\tilde f_h=Z_{h-1}[{C_h^{-1}\over
Z_{h-1}(k')}-{C_{h-1}^{-1}\over Z_{h-1}}]$.
After this integration $\NN$ is, up to a constant, of the form of eq.(3)
with $h$ replaced by $h-1$,
and we can iterate.
{\it In other words we have to perform a Bogolubov transformation
for each scale, as the "mass" $\s_h$ has a non trivial RG flow and it is
different for any $h$; at the same time one has to take into account
the wave function renormalization $Z_h$}.
Let be
$h^*={\rm inf_h}\{\g^{h}\geq |\s_h|\}$.
Note that, if $h^*$ is finite uniformly in $N,\b$ so that
$|\s_{h^*-1}|\g^{-h^*+1}\ge 1$ then
$|g^{< h^*}(\xx)|\le {1\over Z_{h^*}}
{C_M\g^{h^*}\over 1+(\g^{h^*}|\xx|)^M}$;
moreover if $h\ge h^*$ we have $|g^h_{\o,\o}(\xx)|\le
{1\over Z_{h}}
{C_M\g^{h}\over 1+(\g^{h}|\xx|)^M}$ and
$|g^h_{\o,-\o}(\xx)|\le
{1\over Z_{h}}{|\s_h|\over \g^h}
{C_M\g^{h}\over 1+(\g^{h}|\xx|)^M}$.
{\it The propagator for the integration of all the scales
$h< h^*$ obeys to the same bound of a single scale propagator for
$h\ge h^*$. Moreover for $h\ge h^*$ the bound for the non diagonal
propagator
has a factor more ${|\s_h|\over \g^h}$ with respect to the diagonal
propagator}. Finally $g^h_{\oo,\oo}(x-y)=g^h_{\oo,L}(x-y)+C^h_{1,\oo}(x-y)+
C^h_{2,\oo}(x-y)$, with
$g^h_{\oo,L}(x-y)$ given by
$\int d\kk' {e^{i\kk'x}\over {Z_h}}
\frac{f(\g^{-2h}(k_0^2+k^{\prime 2}))}{-ik_0-
2\pi\oo k'}$
which is just the propagator ``at scale h'' of
the Luttinger model, and the other two terms verify the bound of
$g^h_{\oo,\oo}(\xx;\yy)$ with an extra factor $\g^h$ or
${|\sigma_h|\over \g^h}$.

{\rm Lemma}{\it  Assume that $h^*$ is
is finite uniformly in $N,\b$ and
that for any $h>k\ge h^*$
there exists an $\e$ such that $|\vec v_h|\le\e$
and $|{\s_{h+1}\over\s_{h}}|\le \g^{c_a\e}$,
$|{Z_{h+1}\over Z_{h}}|\le \g^{c_b\e^2}$ with $c_a,c_b$
postive constants.
Then $V^k$ is given by a convergent series.}

We have to show by the study of the beta function that
the conditions in the lemma
are verified.
It is possible to choose [8],[9]
the counterterm $\nu$
so that $|\nu_h|<\e$ for all $0\geq h\geq
h^*$.
The Beta function can be written, for $0\geq h\geq
h^*$. :
\begin{eqnarray}\label{A2}
&&\l_{h-1}=\l_h+G^{1,h}_\l+G^{2,h}_{\l}+\g^h R^{h}_\l \nn\\
&&\sigma_{h-1}=\sigma_h
+G_{\sigma}^{1,h}\nn\\
&&\d_{h-1}=\d_h+G^{1,h}_\d
+G^{2,h}_{\d}+\g^h R^{h}_\d\nn\\
&&\frac{Z_{h-1}}{Z_h}=1+G_{z}^{1,h}+
G^{2,h}_{z}+\g^hR^{h}_z
\end{eqnarray}
where a) $G^{1,h}_\l$,
$G^{1,h}_\d$
and
$G^{1,h}_z$ depend only on $\l_0,\d_0;...\l_h,\d_h$ and
are given by series
of terms involving only the Luttinger model part of the propagator
$g^k_{\oo,L}(x-y)$, so they coincide with the Luttinger model Beta
function [8],[9]; b)
$G^{1,h}_\sigma$, $G^{2,h}_\l,G^{2,h}_\d, G^{2,h}_z$
are given by a series
of terms involving at least a propagator $C_{2,\oo}^k(x-y)$ or
$g^{k}_{\oo,-\oo}(x-y)$ with $k\geq h$; c) $R^{h}_i$, $i=\l,z,\d$
are given by a series
of terms involving at least a propagator $C_{1,\oo}^k(x-y)$, $k\geq
h$.
By a simple computation
$G^{1,h}_z=\l_h^2[\b_1+\bar G_z^h]$,
$G^{1,h}_\sigma=\l_h\s_h[-\b_2 +\bar G^h_\s]$,
with $\b_1,\b_2>0$ and $\bar G_z^h,\bar G^h_\s=O(\l_h)$.
{\it Moreover $G^{1,h}_\l$, $G^{1,h}_\d$ coincide by definition with the
Luttinger model Beta function, and it was proved in [8],[9]
that it is vanishing at any order}, {\it i.e.}
$G^{1,h}_\l(\l,\d;\ldots;\l,\d)=0$ and
$G^{1,h}_\d(\l,\d;\ldots;\l,\d)=0$.
Finally as
$|G_\l^{2,h}|,|G_\d^{2,h}|,|G_z^{2,h}|\leq K\e^2 |\s_h|\g^{-h}$,
one finds, for $h\ge h^*$,
$|\l_{h-1}-\l_{0}|<c_1 l_0^2$, $|\d_{h-1}-\d_0|\le c_1 l_0^2$,
$$l_0 \b_1c_2
h\leq\log\left(\frac{\sigma_{h-1}}
{\sigma_0}\right)\leq l_0\b_1 c_3 h$$
$$-\b_3 c_4 l_0^2h\leq \log(Z_{h-1})\leq-\b_3 c_5 l_0^2h$$
for suitable positive constants $c_i$,
\hbox{\it i.e.\ } as usual in models to which the RG is succesfully
applied the flow is essentially {\it described
by the second order truncation of the
beta function}. This shows that it is possible to choose $J_3$ so small
that the conditions of the above lemma are fulfilled.
We call $\h_1=-\log Z_{h^*}/\log u$, $1+\h_2=\log \s_{h^*}/\log u$.

Finally as we said
the integrations of the
$\psi^{(< h^*)}$ is essentially {\it equivalent to the integration
of a single scale $h\ge h^*$}, so it is well defined
by the preceding
arguments.

It is a standard matter to deduce an expansion for the correlations
from the effective potential, and so deducing the results for
$S^{\s_1,\s_2}(\xx)$. Moreover we call $\s(k), Z(k)$ respectively $\s_h, Z_h$ for
$\g^h\le |k|\le \g^{h+1}$ for $h\ge h^*$ and $\s_{h^*}, Z_{h^*}$
for $|k|\le \g^{h^*}$.

Finally we discuss
the modifications in the proof of the above lemma with respect to the
one existing in literature for similar models,[8],[9], mainly
due to the fact that
$[\NN,H]\not=0$. $V^k$
can be written as a sum over "Feynman
graphs" obtained in following way.
Let us consider $n$ points and enclose them
into a set of {\it clusters} $v$ to which
a scale $h_v$ is associated; an inclusion relation is established between
the clusters, in such a way that the innermost clusters are the clusters with
highest scale, {\it i.e.}
if $v'$ is the cluster
containing $v$ then $h_{v'}<h_v$; $v_0$ is the largest cluster.
A set of clusters can be represented
as a {\it
tree}
and the set of the
possible trees
is denoted by $\t_{n,k}$.
To each point contained in a cluster $v$ but not in any smaller
one we associate one of the elements of $\LL V^{h_v}$,
expressed graphically as
a vertex with $2$ or $4$ "half lines".
The half lines are paired in all the possible compatible way and
to the paired lines
we associate a propagator $g^{h_v}_{\oo,\oo'}$
if the paired line is enclosed in the cluster $v$ but not in any
smaller one. The indices of the external lines
of $v$ are denoted by $P_v$
and their number by $|P_v|$.
The $\RR=1-\LL$ operation
acts on the cluster and its action, as we
know,
depend on $P_v$. We call $V^k_n$ the contribution to $V^k$
from $\t\in \t_{n,k}$, and
$|V^k_n| (L\b\g^{-k D(P_{v_0})})^{-1}\equiv |\bar V_n^k|$ is bounded by, if $C_1$
is a constant
$$C_1^n\e^n\sum_{\t\in \t_{n,k}}[\sum_{\{P_v\}}
\prod_{v}
\g^{-[D(P_v)+z(P_v)](h_v-h_{v'})}]$$
$$[\prod_{v\in V_2}
|\sigma_{h_v}|\g^{-h_v}]$$
where $D(P_v)=-2+|P_v|/2$,
and $V_2$ is the set of the clusters $v$ containing a non
diagonal propagator $g^{h_v}_{\o,-\o}$ {\it i.e.} the smallest
clusters containing a non diagonal propagator.
The factor $z(P_v)$ (which would be zero if we had naively taken $\RR=1$)
is defined as:1)
$z(P_v)=1$ if $|P_v|=4$ and $\d_4^a=1$; 2)
$z(P_v)=2$ if $|P_v|=2$ and $\d_2^a=1$; 3)
$z(P_v)=1$ if $|P_v|=2$ and $\d_2^b=1$; 4)$z(P_v)=0$
in the remaining cases. The first parenthesis is the
power
counting of a graph, and the second is due to the extra factor
in the
bound for the non diagonal propagators; the only non trivial part of the
above bound is the use of the Gramm-Hadamard inequality to take into
account the relative signs of the graphs (estimating them
by their absolute value one would get
an extra $k!$ in the bounds spoling convergence).
If $D(P_v)+z(P_v)>0$
the sum over $\t,\{P_v\}$
can be bounded and $|\bar V_n^k|\le C_2^n \e^n$; however, if
$\d_4^a,\d_2^a,\d_2^b=0$
this is not the case. The factors $\d_4^a,\d_2^a,\d_2^b$ are products
of two deltas. If the first delta is vanishing, by the
support in momentum space of $g^h_{\o,\o'}$,
it follows that there exists a fixed scale $\bar h$, indipendent on
$L,\b,k,n$, such that there are clusters $v$ with the first non vanishing
deltas only if $h_v\ge \bar h$; so these clusters give no problems
(one can even choose the functions $\chi$ so that $\bar h\equiv 0$).
Let us consider now the case in which the second deltas are non vanishing.
Note that
$|\sigma_{h_v}\g^{-h_v}|\le |
{\sigma_{h_v}\over \sigma_{h^*}}{\sigma_{h^*}
\over \g^{h_v}}|$ $\le \g^{(k-h_v)(1-O(\e))}\le
\g^{(k-h_v)(1/2)} $
as $|\s_{h^*}|\le \g^{h_*}$.

Let us consider vertices $v_1,v_2,..,v_I$
ordered so that  $v_1>v_2>...>v_I$ ; then there exists at
least a
non diagonal propagator with scale $h_{v_1}$ and
$$|\sigma_{h_{v_1}}\g^{-h_{v_1}}|\le
\g^{(k-h_{v_1})(1/2)}\le
\g^{(h_{v'_I}-h_{v_I})(1/2)}$$
$$ \g^{(h_{v_I}-h_{v_1})(1/2)}\leq
\g^{(h_{v'_I}-h_{v_I})(1/2)} \g^{(h_{v'_{I-1}}-h_{v_1})(1/2)}...$$
so that
$\prod_{v\in V_2}
|\sigma_{h_v}|\g^{-h_v}\le
\prod_{v \in V_b}
\g^{-(1/2)(h_v-h_{v'})}$, if $V_b$ are the clusters not verifing the
second
deltas.

At the end
$$[\prod_{v}
\g^{-[D(P_v)+z(P_v)](h_v-h_{v'})}][\prod_{v\in V_2}
|\sigma_{h_v}|\g^{-h_v}]\le$$
$$[\prod_{v}
\g^{-(1/2)[D(P_v)+\bar z(P_v)](h_v-h_{v'})}]$$
with $D(P_v)+\bar z(P_v)> 0$, for any $v$. Hence a
$C^n\e^n$ bound follows for $|\bar V^n_k|$.

{\bf Acknowledgments} I thank G.Benfatto and G.Gallavotti for many
important discussions.

\section{\bf References}

\begin{description}

\item[[1]] R.J.Baxter. Phys. Rev. Lett. {\bf 26}, 832 (1971);
J.D.Johnson, S. Krinsky, B.M.McCoy. Phys. Rev. A. {\bf
                8},5,2526-2547 (1973)
\item[[2]] B.Suterland. Journ. Math. Phys.
                {\bf 11},11,3183-3186, (1970)
\item[[3]] L.Ko, B.M.McCoy, Phys.Rev.Lett. {\bf 56}, 24, 2645-2648 (1986)

\item[[4]] A.Luther, I. Peschel, Phys. Rev. B {\bf 12},
                  3908--3917 (1975); E.Fradkin, Field Theoris of
Condensed matter systems, Add. Wesley (1991)

\item[[5]] V. Dotsenko, Jour. Stat.Phys. {\bf 34}, 781 (1984)

\item[[6]] G.Benfatto, G.Gallavotti, Renormalization Group, Princeton
paperbacks (1995)

\item[[7]] E.Lieb, T. Schultz, D. Mattis, Ann. of. Phys. {\bf 16},
407-466 (1961); B.M. McCoy, Phys. Rev. {\bf 173}, 2,531-541 (1968)

\item[[8]] G. Benfatto, G. Gallavotti, A. Procacci,
                  B. Scoppola, Comm. Math. Phys {\bf 160}, 93-171
\item[[9]] F. Bonetto, V. Mastropietro,
Comm. Math. Phys. {\bf 172}, 57-93 (1995);
 Math. Phys. Elec. Journ. {\bf 1}
(1996);
Phys. Rev. B 56 1296-1308 (1997); Nucl. Phys. B 497 541-554
(1997);V.Mastropietro, submitted to Comm .Math .Phys.

\end{description}
\end{document}